\documentstyle[twoside]{article}
\input psfig

\begin{document}

\def\apj #1 {{ApJ} #1, }
\def\pasp #1 {{PASP} #1, }
\def\lin{\vskip12pt\indent}
\def\reef{\hangindent=20pt\hangafter=1}
\def\apjl #1 {{ApJL} #1, }
\def\apjs #1 {{ApJS} #1, }
\def\apss #1 {{A\&SS} #1, }
\def\exa  #1 {{Exp Astr} #1, }
\def\aa #1 {{A\&A} #1, }
\def\aar #1 {{A\&AR} #1, }
\def\araa #1 {{ARA\&A} #1, }
\def\aas #1 {{A\&AS} #1, }
\def\mn #1 {{MNRAS} #1, }
\def\aj #1 {{AJ} #1, }
\def\qjras #1 {{QJRAS} #1, }
\def\mic{\,\mu \rm m}                
\def\ms{M$_{\odot}$}
\def\msb{M$_{\odot}$~}

\def\la{\mathrel{\mathchoice {\vcenter{\offinterlineskip\halign{\hfil
$\displaystyle##$\hfil\cr<\cr\noalign{\vskip1.5pt}\sim\cr}}}
{\vcenter{\offinterlineskip\halign{\hfil$\textstyle##$\hfil\cr<\cr
\noalign{\vskip1.0pt}\sim\cr}}}
{\vcenter{\offinterlineskip\halign{\hfil$\scriptstyle##$\hfil\cr<\cr
\noalign{\vskip0.5pt}\sim\cr}}}
{\vcenter{\offinterlineskip\halign{\hfil$\scriptscriptstyle##$\hfil
\cr<\cr\noalign{\vskip0.5pt}\sim\cr}}}}}

\def\ga{\mathrel{\mathchoice {\vcenter{\offinterlineskip\halign{\hfil
$\displaystyle##$\hfil\cr>\cr\noalign{\vskip1.5pt}\sim\cr}}}
{\vcenter{\offinterlineskip\halign{\hfil$\textstyle##$\hfil\cr>\cr
\noalign{\vskip1.0pt}\sim\cr}}}
{\vcenter{\offinterlineskip\halign{\hfil$\scriptstyle##$\hfil\cr>\cr
\noalign{\vskip0.5pt}\sim\cr}}}
{\vcenter{\offinterlineskip\halign{\hfil$\scriptscriptstyle##$\hfil
\cr<\cr\noalign{\vskip0.5pt}\sim\cr}}}}}


\title{Mid-infrared imaging of AGB star envelopes.\\
II. Modelling of observed sources}

\author{M. Marengo$^*$\\ 
{\small SISSA/ISAS - Trieste (Italy)}\\
G. Canil, G.Silvestro\\
{\small Dipartimento di Fisica Generale - Universit\`a di Torino (Italy)}\\
L. Origlia, M. Busso\\
{\small Osservatorio Astronomico di Torino, Pino Torinese (Italy)}\\
P. Persi\\ 
{\small Istituto di Astrofisica Spaziale, CNR, Frascati (Italy)}}

\date{}

\maketitle

\begin{abstract}
Radiative transfer modelling of AGB circumstellar envelopes
is applied to a sample of AGB stars previously observed with the
mid-IR imaging camera TIRCAM (Busso et al. 1996: Paper~I). We present
the results of our simulations, aimed at  
deriving the physical parameters of the envelope, such as the
optical depth and the radial thermal structure,  
the mass loss and the dust-to-gas mass ratio. The
chemical composition of the dust in the observed envelopes is
discussed. The ability of 
different sets of dust opacities to fit the mid-infrared spectra 
is evaluated. The hypothesis of dust grain 
aging and annealing in O-rich envelopes is considered in order 
to explain an apparent inadequacy of the
availabe opacities to describe the variety of observed spectra,
as previously noted by other authors. Various possible origins of
the discrepancies are discussed, together with their consequences 
on the dust grain formation processes.
\end{abstract}

\begin{center}
SISSA Ref: 113/96/A (July 1996)\\\
Submitted to Astronomy and Astrophysics
\end{center}

\vfill
\hrule
\vspace{0.3cm}
$^*$e-mail: marengo@sissa.it

\newpage




\section{Introduction}

Asymptotic Giant Branch (AGB) stars are characterized by intense mass
loss processes that play a central role in the star's subsequent
evolution and
lead to the formation of a cold circumstellar envelope
(CSE) of gas and dust.
The envelope chemical composition is determined by the amount
of  
carbon enrichment in the stellar atmosphere, due to mixing processes
(the so called {\it third dredge-up}), induced by thermal pulses 
near the end of the AGB phase.
C-rich envelopes (having [C]/[O]$>$1 by number) mainly exist around carbon
stars, while O-rich CSEs ([C]/[O]$<$1) are associated to M
giants. The transition from O-rich to C-rich envelopes is controlled by the
extent of the third dredge-up, and it can occur only for AGB stars 
in a narrow mass range: 
indeed, stars with a core mass lower than 0.6 \msb (Straniero et
al. 1995) will not undergo the third dredge-up, while H burning at the base
of the convective envelope (Hot Bottom Burning) is expected to 
deplete carbon in
the envelope of stars more massive than $\sim$ 5 \msb
(Wood et al. 1983; Boothroyd et al. 1993; Frost \& Lattanzio
1995), preventing them from becoming C-rich.

Despite the low dust-to-gas mass ratio 
(in the range $\sim$0.001-0.01), the optical
properties of AGB CSEs are mainly determined by the dust grains:
silicates in O-rich envelopes (P\'egouri\'e \& Papoular 1985) and a
mixture of Hydrogenated Amorphous Carbon (HAC, Jones et al. 1990; 
Duley 1993) with inclusions of SiC  (Skinner \& Whitmore 1988)
and possibly Policyclic Aromatic Hydrocarbons (PAH, Puget \& L\'eger,
1989; Cherchneff \& Barker 1992) in C-rich CSEs. 
Both oxidic and carbonaceous star dust is characterized by vibrational
bands positioned in the mid-IR window (at 9.7 and 18 $\mu$m for
silicates, at 11.3 $\mu$m for SiC and at 3.3, 6.2, 7.7, 8.6 and 11.3
$\mu$m for PAH); furthermore, most of the continous thermal
radiation from the dusty optically opaque envelope is
emitted in the same wavelenght range.

For these reasons a mid-IR search was carried out 
with the imaging camera TIRCAM (Busso et
al. 1996, hereafter Paper~I) for a sample of 16 sources, including
both AGB (O-rich and C-rich) and post-AGB stars.
In Paper~I, photometry and colors were derived using 10\% bandwidth
filters at 8.8, 9.8, 11.7 and 12.5 $\mu$m, and compared with the Low
Resolution Spectra (LRS) from IRAS (1986), in order to establish 
suitable photometric criteria for discriminating between O-rich and
C-rich sources and for estimating mass loss rates. The observations
formed the basis for discussing the evolutionary status of the sources
and their mass loss history. 
In Section~2 we discuss
the general features of their spectra, from the point of
view of the optical properties that can be inferred for dust.

Detailed numerical simulations of the CSE thermal structure is
the subsequent step for a better determination of the physical
parameters. One has to compute the source spectrum and the radial
brightness distribution for a wide range of input parameters;
comparison with observations will then
allow us to estimate the optical depth of the envelopes, to derive
mass loss rates, and to extract information on chemical abundances
which are
relevant to study the stellar nucleosynthesis (see e.g. Busso et al. 
1995). 
With these aims in mind, we developed a numerical code for solving the
problem of non-grey radiative transfer through a spherically symmetric
dust shell around an evolved AGB star. We present our model in
Section~3, with emphasis on the input model parameters
and on the 
dust opacities used. In Section~4 the model results are shown for the
sources observed in Paper~I whose LRS is known, and a new estimate for the
dust mass loss rates and the dust-to-gas mass ratio is obtained. 

In Section~5 a more general discussion of the model spectra is
carried on. We point out how available dust opacities seem not to 
be
completely adequate for a detailed simulation of the mid-IR features
of circumstellar dust, as previously suggested by different authors
(Little-Marenin \& Little 1990, hereafter LML90; Simpson 1991), 
and we test possible explanations. 
Our conclusions are summarized in Section~6.


\section{Spectra of AGB Circumstellar Envelopes}

The IRAS color-color diagram (van der Veen \& Habing 1988) obtained
by IR photometry at 12, 25 and 60 $\mu$m,  provides a first-order
discrimination of AGB sources, according to their infrared excess,
variability and C abundance. However, as shown in Paper~I, 
such IRAS color-color
diagram can lead to misclassification of AGB CSEs due to the large
width of IRAS filters (this is true in particular for the one at 
12 $\mu$m, which cannot
discriminate between silicates and carbonaceous features). 
 Although narrow band filters allow a better discrimination 
(as in the case of
TIRCAM photometric system), a detailed description of the dust
emission/absorption features requires mid-IR spectra. 

IRAS LRS (1986) provide spectral energy distributions in
the wavelength range 7.7-22.6 $\mu$m for about 4,000 AGB sources, which 
are classified with a two-digit code. 
The first digit (main class) is related to the spectral index
$\beta$ of the continuum (assuming $F_\lambda \propto
\lambda^\beta$),  while the second 
indicates the strength of the main
feature, in emission or absorption, present in the mid-IR
window. In particular, spectra of O-rich sources are identified by
the 9.7 $\mu$m silicate feature, that is found in emission or 
absorption in the LRS classes 2$n$ and 3$n$, respectively. 
C-rich envelopes are mainly found in the 4$n$ class, defined by the
presence of the 11.3 $\mu$m SiC feature in emission. Post-AGB objects
of both types are in the main classes 6$n$ and 7$n$, characterized by
very large infrared excess. Finally, both O-rich and C-rich sources
may fall in classes 1$n$ and 0$n$ (featureless spectra), where a few
LRS with self-absorbed SiC feature can be found (Omont et al. 1993).

Table~1 lists the spectral characteristics of the sources in
Paper~I with known LRS; the star AFGL 1822 was added to the
TIRCAM sample in order to have an evolved O-rich envelope with the 9.7
$\mu$m silicate feature in absorption. Post-AGB sources have not been
included because their modelling involves very different physical 
conditions, dominated by the hot radiation emitted by the
central star (B or A spectral type) on its way to become a white dwarf.  

\begin{table}

\begin{center}
\begin{tabular}{||r|c|l|c|c|l|c||} \hline
\# & IRAS name & Name & Type & SP type & LRS & LRS sub. \\ \hline \hline
 1 & 02143+4404 & W And & S-star & S8,2e(M) & 22 & S \\ \hline
 2 & 08525+1725 & X CnC & C-rich & C5,4(SRb) & 42 & ---  \\ \hline
 3 & 09425+3444 & R LMi & O-rich & M7e(M) & 24 & Sil+  \\ \hline
 4 & 09452+1330 & CW Leo & C-rich & Ce(M) & 43 & --- \\ \hline 
 5 & 12427+4542 & Y CVn & C-rich & C5,5J(SRb) & 42 & ---  \\ \hline
 6 & 13001+0527 & RT Vir & O-rich & M8III(SRb) & 21 & --- \\ \hline 
 7 & 15193+3132 & S CrB & O-rich & M6,5e(M) & 24 & Sil+ \\ \hline 
 8 & 15255+1944 & WX Ser & O-rich & M8e(M) & 29 & Sil \\ \hline 
 9 & 16029-3041 & AFGL 1822 & O-rich & M & 32 & --- \\ \hline 
10 & 16235+1900 & U Her & O-rich & M7IIIe(M) & 23 & Sil+ \\ \hline 
11 & 18040-0941 & FX Ser & C-rich & C(Lb) & 44 & ---  \\ \hline
12 & 18397+1738 & NSV 11225 & C-rich & Ce & 43 & ---  \\ \hline
\end{tabular}
\end{center}

\caption[1]{List of AGB sources from Paper~I with known LRS. The
evolved O-rich envelope AFGL 1822 was added to have a 3$n$ LRS class
source. Last column: LML90 LRS subclass, when available.}
\end{table}

Note that the LRS
class 2$n$ contains sources with different characteristics; 
a tentative classification is in the  
LML90 paper, where the silicate band
profiles in optically thin O-rich envelopes are divided into seven
subclasses, according to their shape and the presence of secondary
peaks.
The subclass for our O-rich sources, when given by LML90, is 
reported in Table~1; the S-star W And is tentatively
attributed to the ``S'' feature subclass, characterized by a broad peak at
10.3 $\mu$m. The other sources fall in subclass ``Sil'' (with a strong
feature peaked at 9.8 $\mu$m, typical of 2$n$ LRS with $n>4$) and
subclass ``Sil+'' (having a secondary ``bump'' at 11.3 $\mu$m).
LML90, and Simpson (1991) suggest that these
differences be related to a kind of ``mineralogical diversity''
in the oxidic CSE dust, although this explanation is still
controversial (see e.g. Ivezi\'c \& Elitzur 1995, hereafter IE95). 
We tried to model
the shape of the silicate feature in our sources belonging to different
subclasses using several sets of opacities and dust mixtures.


\section{Our Model}

Radiative transfer in circumstellar dust shells has been previously
discussed by many authors (Rowan-Robinson 1980; Griffin 1990;
Justtanont \& Tielens 1992, Groenewegen, 1993
and, more recently, Hashimoto 1995; a detailed discussion of these
models is given in Habing, 1996). 
A fully self-consistent modelling of
envelopes around evolved late-type stars requires coupling of
radiative transfer with the hydrodynamic equations of motion for the two
interacting fluids (gas and dust) which constitute the stellar
wind. Such an approach has been attempted by IE95: 
their treatment, in the hypothesis of a steady-state outflow
due to constant dust-driven stellar wind, can successfully account for
the observed mass loss rates and is able to reproduce the IRAS colors
and visibility functions (see Ivezi\'c and Elitzur 1996a, b) 
for a large sample of AGB objects in all LRS classes. More recently,
models taking into account time dependent outflows have been developed
to study the effect of sudden variations in the stellar parameters in
connection with thermal pulses (Sh\"onberner 1996). 
This analysis is
necessary in order to follow the latest stages of AGB evolution, which
lead to the formation of a planetary nebula; however, time variations 
of the mass loss rate, on a thermal pulse time scale, have negligible
effects on the IR properties of at least 95\% of late-type stars
associated with IRAS sources (IE95).

In our model we totally neglect the dynamics of the envelope; rather,
we concentrate our efforts on obtaining an accurate fit of the
observed sources, paying special attention to the mid-IR interval,
where the TIRCAM spectral bands are located (see Paper I).  The model
parameters are fixed by fitting the IRAS and TIRCAM data. 
For any dust mixture, the spectrum (and in
particular the shape of the features) is used to estimate the optical
depth $\tau_\nu$ of the envelope, proportional to the dust opacity and the
mass loss rate. Near- and Mid-IR photometric data allow us to   
determine the stellar parameters $T_{eff}$, $R_*$
(the central star is assumed to radiate a black body spectrum at
temperature $T_{eff}$), and the
inner CSE radius $R_1$; the outer radius $R_2$ is derived by fitting the
IRAS far-IR photometry at 60 and 100 $\mu$m. The source distance enters
in the model only as a scale factor for the flux densities, and is    
determined by normalizing the spectra with the observed photometry. We
first assumed the value estimated by Loup et al. (1993); then we 
computed our own value of $d = \sqrt {L_*/F_{1 kpc}}$, where
$L_* = 4 \pi \sigma R_*^2 T_{eff}^4$ is the total luminosity and
$F_{1 kpc}$ the total flux (in L$_\odot$, assuming a reference 
distance of 1 kpc) obtained by Loup et al. (1993) using IRAS fluxes.  

The code computes iteratively a self-consistent thermal structure of
the envelope; the emergent spectra is 
calculated taking into account the effect of non-isotropic scattering,
absorption and thermal reemission by grains. 
The computation is performed on the
hypotheses of: (i) spherical symmetry of the dust shell, with an $n(r)
\propto r^{-2}$ density distribution, consistent with a steady outflow
at constant velocity, (ii) balance between absorption and emission by
the dust grains, and (iii) LTE dust radiation at the local temperature
$T(r)$. 

The dust grain composition is simulated adopting different sets
of dust opacities $Q_\nu$; the dust grain size is usually assumed to 
satisfy a size  
distribution $n(a) \propto a^{-3.5}$ with 0.01 $\mu$m $\la a \la$ 0.25
$\mu$m (Mathis, Rumpl and Nordsieck 1977, hereafter MRN; see
e.g. Jura, 1996 for more recent data).  
The main effect of considering a 
grain size distribution is a broadening of the 9.8
$\mu$m silicate feature (Simpson 1991), due to the non 
linear dependance of the opacity on $a$; this is however important
only for $\lambda \la a$ and in the case of AGB CSE grains can be
neglected. Hence we simply adopted for the grain size 
the average MRN value $a$ =
0.1 $\mu$m.


\subsection{Dust Opacities}

Following a Chan \& Kwok (1990) scaling theorem, 
the solution of the radiative transfer equation
in spherical symmetry is scaled on the flux-averaged optical depth
$\tau_F = \int_0^\infty \tau_\nu F_\nu d\nu / \int_0^\infty F_\nu
d\nu$, where:

$$\tau_\nu = \pi a^2 Q_\nu \int_{R_1}^\infty n(r) dr \eqno (1)$$

In equation~1, $\tau_\nu$ depends on frequency only through
the opacity $Q_\nu$, which is thus the main parameter for 
modelling the spectra; for this reason, a
great part of our preliminary work consisted in testing the various
opacity profiles available in the literature. We now present the results
of our analysis, separately for O-rich and C-rich dust.

It is well known that O-rich dust contains essentially silicate
grains: they can be present as piroxens (Mg,Fe)SiO$_3$, or as
olivin\ae \ ((Mg,Fe)$_2$SiO$_4$). 
Actually, the olivina
opacity profile of Kr\"atschmer \& Huffmann (1979) is suitable for
supergiants, but not for Miras like WX Serpentis (see Griffin 1993).
In the case of circumstellar envelopes the silicate dust probably
consist of piroxens only, 
with various types of impurities making them ``dirty silicates''. We
considered five opacity profiles for astronomical silicates.

The first is the one by Jones \& Merril (1976); they introduced 
the concept of ``dirty silicates'', 
though many details of their opacity profile were subsequently 
modified.

The second is by Draine \& Lee (1984); despite its wide use and 
accuracy, this is not well suited for our needs,
since it is based on observations of the interstellar medium, 
and not of the circumstellar one (see also Ossenkopf et al. 1992, 
hereafter OS92, and Griffin 1993). 

The third is by Volk \& Kwok (1988): it is based
on the observed spectra of 467 AGB stars with good quality photometric
data in all four IRAS bands; the spectrum obtained from this opacity is
similar to the observed one, but the 9.8 $\mu$m feature is slightly too
much asymmetric and its peak is shifted towards redder wavelengths.  

The fourth profile is the one by OS92: it is
based on the previous but it is chemically and physically consistent,
in the sense that computes the complex dielectric function for the
dust using the Kramers-Kr\"onig relations along with the Mie theory.
Taking into account the possibility of dust annealing in the CSE, as
illustrated by Stencel et al. (1990) and Nuth and Hecht (1990), the
authors give the opacity in two versions, one for ``warm'' silicates
(hereafter Oss1) and the other for ``cold'' ones (hereafter Oss2), 
depending on the physical conditions at their condensation. 

The last and more recent is by David \& P\'egouri\'e (1995), 
and is computed using Mie theory for a set of more than 300 IRAS
sources with known LRS.

We obtained  the best 
results when fitting our AGB O-rich sources   
with OS92 Oss1 and Oss2 opacities.  

In the case of C-rich envelopes, the continuum in the spectra is
usually ascribed to the presence of graphite or amourphous carbon,
while the characteristic band at 11.3 $\mu$m is commonly attributed to
SiC, although the presence of PAH has been proposed as well.   
Graphite is now commonly ruled out, because
of the absence in the observed spectra of its typical narrow band at 11.52
$\mu$m (Draine 1984); moreover, the whole
shape of the spectra obtained by using graphite (the most used
opacities are by Draine \& Lee 1984) does not fit the observed ones,
as we tested directly. Therefore the main constituent of CSE
carbonaceous dust seems to be amourphous carbon: 
due to the variability in its physical and optical properties, a
comparison with observational data is fundamental.
An accurate enough opacity profile, observationally tested for
amorphous carbon is the one by Martin \& Rogers (1987). 
More recently Rouleau 
\& Martin (1991) presented a detailed study giving accurate sets of
optical constants derived by Bussoletti et al. (1987): the set named
AC1, in particular, is similar to the Martin \& Rogers (1987) opacity, 
although more detailed; this is the opacity we used for our simulations.

Finally, in order to obtain spectra with the characteristic 11.3 $\mu$m band,
we admitted the presence of SiC mixed with the amorphous carbon: we
choose the opacity set for $\alpha$-SiC (the exagonal rhomboedric
cristalline form of SiC) from P\'egouri\'e (1988), which at the moment
is the only accurate profile available in the literature; to obtain
the optical properties for the mixture, we simply made a weighted average
of the two species: we never found necessary to take a SiC percentage
greater than 10\%. This strongly depends on the opacity chosen for
amorphous carbon; IE95, for example, used a more
absorptive kind of amorphous carbon, together with P\'egouri\'e (1988) data
for SiC, and found a SiC percentage up to 30\%.  

In Figure~1 we plot the final opacity profiles for silicates, amorphous 
carbon and SiC used in our simulations. 

\begin{figure}[t]
\centerline{\psfig{file=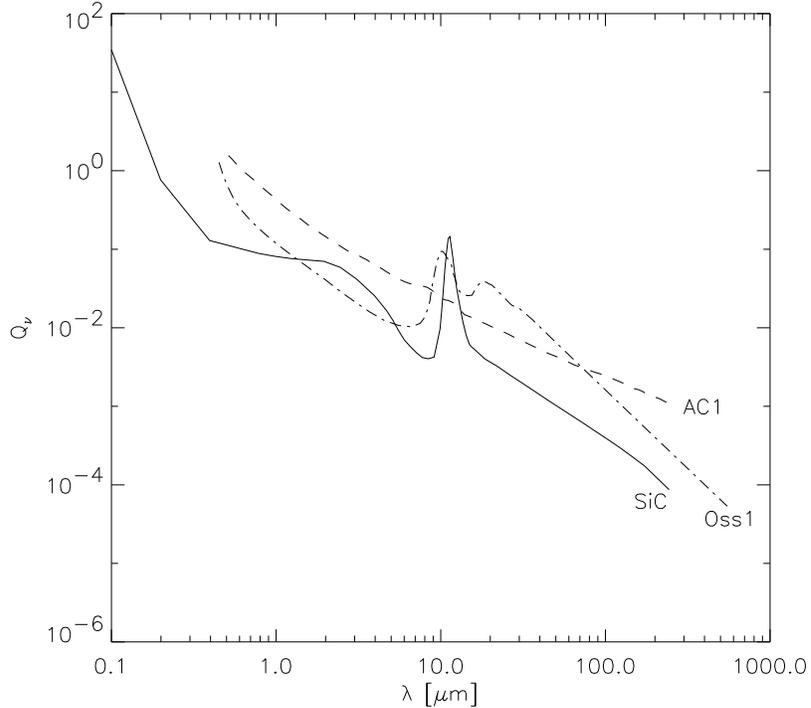,height=10cm}}
\caption[1]{Dust opacities used in our computations:  Ossenkopf~1 ``dirty 
silicates'' profile by OS92 (dot-dash-dot line),
amourphous carbon AC1 by Rouleau \& Martin 1991 (dashed line) and
P\'egouri\'e (1988) SiC (dotted line).}
\end{figure}

In our simulations we have not introduced Polycyclic Aromatic
Hydrocarbons (PAH), because they are mainly excited by UV radiation,
and their contribution to the dust heating and the final spectra is
low for AGB stars, at least if a hot star companion is not present (Buss
et al. 1991).


\section{Model Results}

In order to fit the IR spectra of the
sources in our sample,   
for each set of dust opacities we computed a grid of models parameterized   
on the 10 $\mu $m optical depth $\tau _{10}$,    
which is proportional to the dust mass loss rate of the envelope:

$$\dot M_d = {{16 \pi} \over 3} {{\rho_d a} \over {Q_{10}}}
\tau_{10} R_1 v_e \eqno (2)$$

\noindent
where $\rho_d$ is the density of the
dust grains (we assumed 3.0 g/cm$^3$ for OS92 
silicates, 1.85
g/cm$^3$ for AC1 amorphous carbon and 2.5 g/cm$^3$ for P\'egouri\'e 1988   
SiC), $Q_{10}$ is the opacity at 10 $\mu$m, and $v_e$ is the outflow
velocity of the dust. The models were computed assuming $d$ = 1 kpc,
$R_*$ = $10^{13}$ cm, $T_*$ = 2500 K, $R_2$ = 2000 $R_1$, $R_1$ = 4
$R_*$ if $\tau_10 < 1$ and $R_1$ = 5 $R_*$ if $\tau_{10} > 1$.

Figure~2 shows model spectra (in the wavelength interval
1-100 $\mu$m) of O-rich envelopes for increasing $\log \tau_{10}$, from
$-3$ to 1; in Figure~3 the sequence of C-rich models   
with $\log \tau_{10}$ from $-3$ to 0 is plotted.  
In both cases greater optical depths
cause larger infrared excesses, and the dust features (silicates 9.8
$\mu$m and SiC 11.3 $\mu$m) change from emission to absorption. We
stopped the C-rich sequence at $\tau_{10} = 3$ because this can be
taken as a limit for this class of sources, considering that SiC
features in absorption are rare in the LRS database (see Omont et al. 
1993).  

\begin{figure}[p]
\centerline{\psfig{file=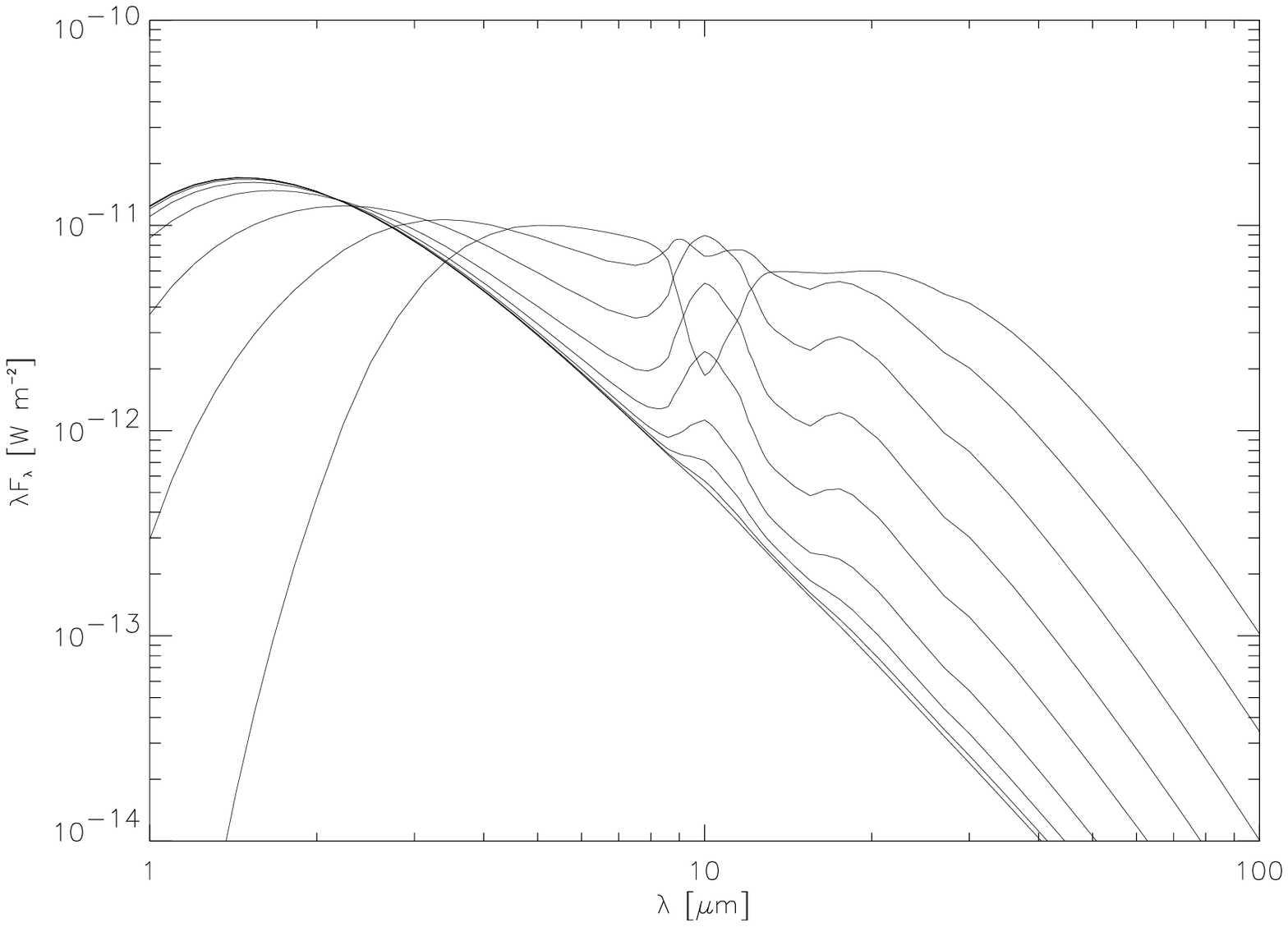,height=8cm}}
\caption[2]{Model spectra for O-rich circumstellar envelopes; the
curves have 
$\tau_{10}$ = 0.001, 0.003, 0.01, 0.03, 0.1, 0.3, 1.0, 3.0 and 10.0;
the IR excess increases with increasing optical depth.}

\centerline{\psfig{file=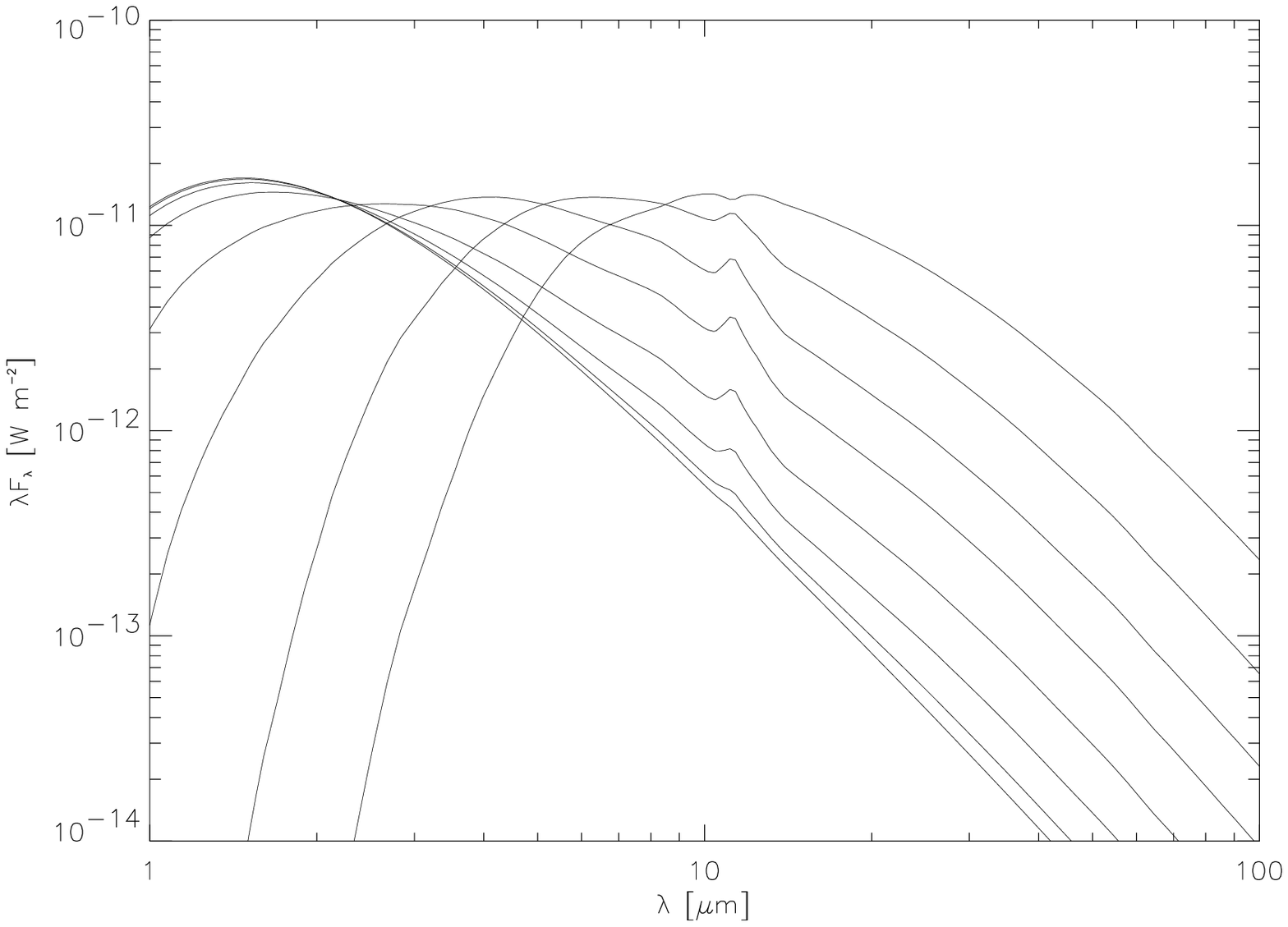,height=8cm}}
\caption[3]{Same as Figure~2, for C-rich circumstellar envelopes having  
$\tau_{10}$ = 0.001, 0.003, 0.01, 0.03, 0.1, 0.3, 1.0 and 3.0.}
\end{figure}

The van der Venn \& Habing (1988) IRAS color-color
diagram for the computed model sequences is shown in Fig.~4. 
The two curves are
associated to increasing optical depth, and thus to increasing mass
loss; note that they cross the appropriate 
regions of the diagram for the two classes of sources.
The curves should not be considered as evolutionary
tracks, because there is indication that mass loss rates do 
not increase
monotonically during AGB evolution, rather they can have sudden variations 
in connection with thermal pulses
(Vassiliadis \& Wood 1993; see also Paper~I).

\begin{figure}
\centerline{\psfig{file=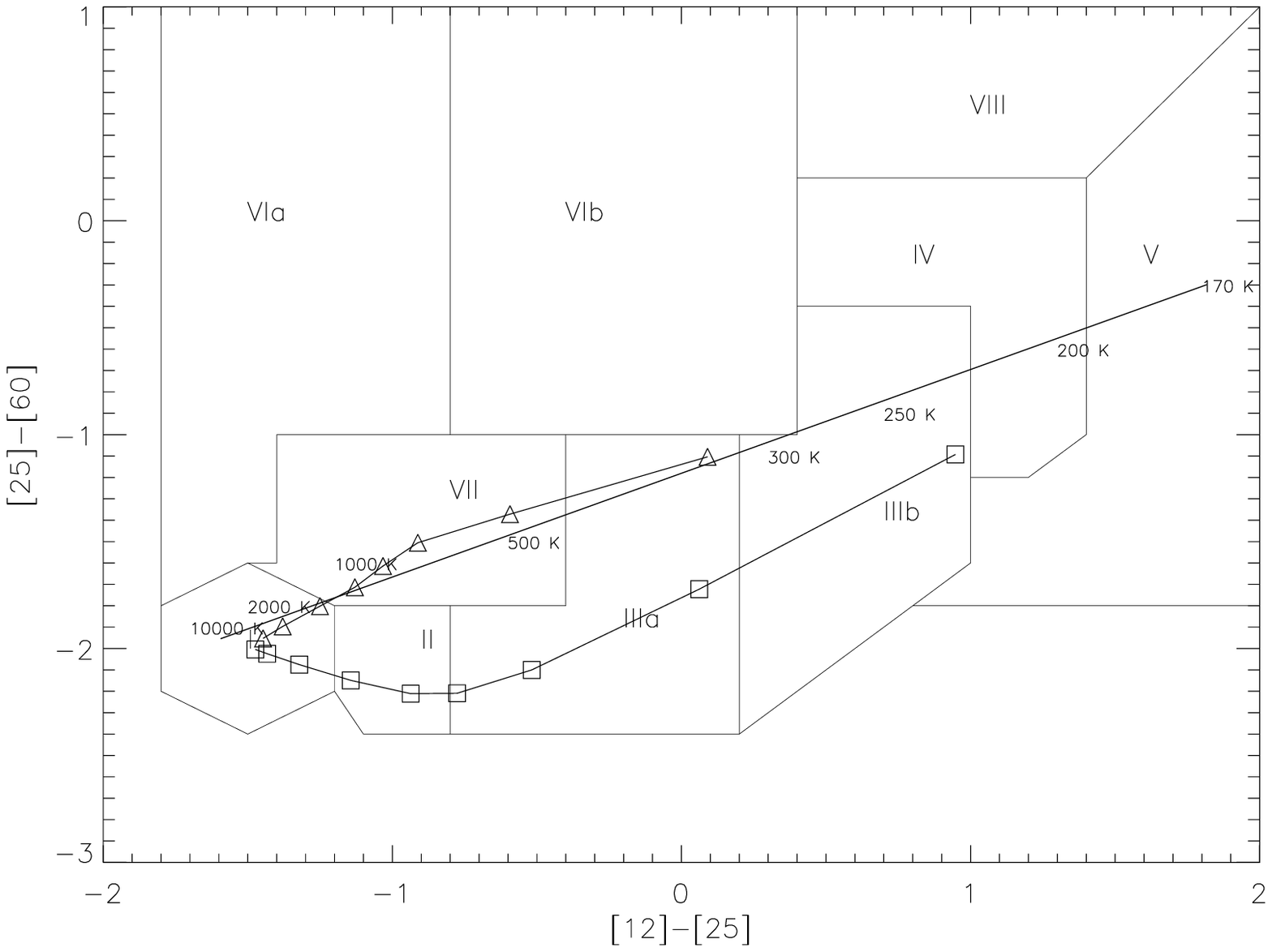,height=13cm}}
\caption[4]{IRAS [25]-[60] vs. [12]-[25] color-color diagram for
O-rich (squares) 
and C-rich (triangles) model in Figure~2 and 4 respectively;
increasing optical depth results in increasing IR excess (from left to
right).} 
\end{figure}


\subsection{Spectral Distribution}

By means of our model we finally fitted the spectral distribution 
of the sources listed in Table~1 and 
we obtained some relevant physical parameters for the star ($T_{eff}$, $R_*$)  
and the envelope ($R_1$, $R_2$, $\tau _{10}$).  

Moreover, we derived the dust mass loss $\dot M_d$ (using Eq.~1, 
with $v_e$ from Loup et al. 1993) 
and the bolometric luminosity $L_{tot} = 4 \pi \sigma R_*^2 T_{eff}^4$. 

From Loup et al. (1993) we also obtained values (based on
radio OH and HCN observations and IRAS data) for the total mass 
loss rate $\dot M$ and the total emitted fluxes $F_{1 kpc} = 
4 \pi (1 \hbox{ kpc})^2 F_{total}$; 
we thus estimated the dust to gas ratio $\mu$ (in mass fraction)
and the distance $d$ for each source. Our results are summarized in
Table~2. 

\begin{table}

\scriptsize
\begin{center}
\begin{tabular}{||r|l|c|c|c|c|c|c|c|c||} \hline
\# & Name & $\tau_{10}$ & $v_e$ & $\dot M$ & $\dot M_d$ & 
$\mu$ & $L_*$ & $F_{1 kpc}$ & $d$ \\ \hline
   &  &  & km/s  & M$_\odot$/yr & M$_\odot$/yr & 
\% & L$_\odot$ & L$_\odot$ & kpc \\ \hline \hline
1 & W And & 0.17 & 20. & $9.7 \cdot 10^{-7}$ & $2.5 \cdot 10^{-9}$ & 
0.26 & $2.9 \cdot 10^4$ & $4.4 \cdot 10^4$ & 0.26 \\ \hline
2 & X CnC & 0.15 & 9.7 & $4.6 \cdot 10^{-7}$ & $1.5 \cdot 10^{-9}$ & 
0.33 & $1.3 \cdot 10^4$ & $2.1 \cdot 10^4$ & 0.79 \\ \hline
3 & R LMi & 0.15 & 7.0 & $5.0 \cdot 10^{-7}$ & $1.0 \cdot 10^{-9}$ & 
0.20 & $5.7 \cdot 10^3$ & $9.4 \cdot 10^4$ & 0.25 \\ \hline
4 & CW Leo & 1.30 & 15. & $4.8 \cdot 10^{-5}$ & $5.0 \cdot 10^{-8}$ & 
0.10 & $4.6 \cdot 10^4$ & $6.6 \cdot 10^5$ & 0.26 \\ \hline
5 & Y CVn & 0.01 & 8.2 & $2.8 \cdot 10^{-7}$ & $1.7 \cdot 10^{-10}$ & 
0.06 & $6.0 \cdot 10^3$ & $8.7 \cdot 10^4$ & 0.26 \\ \hline
6 & RT Vir & 0.15 & 9.3 & $1.3 \cdot 10^{-6}$ & $4.3 \cdot 10^{-9}$ & 
0.33 & $5.9 \cdot 10^4$ & $1.3 \cdot 10^5$ & 0.67 \\ \hline
7 & S CrB & 0.15 & 6.3 & $6.0 \cdot 10^{-7}$ & $8.0 \cdot 10^{-10}$ & 
0.14 & $4.6 \cdot 10^3$ & $2.3 \cdot 10^4$ & 0.45 \\ \hline
8 & WX Ser & 1.30 & 8.8 & $5.3 \cdot 10^{-7}$ & $8.2 \cdot 10^{-9}$ & 
1.55 & $3.2 \cdot 10^3$ & $7.2 \cdot 10^3$ & 0.67 \\ \hline
9 & AFGL 1822 & 8.00 & 17. & $2.5 \cdot 10^{-5}$ & $2.0 \cdot 10^{-8}$ & 
0.08 & $1.2 \cdot 10^4$ & $2.8 \cdot 10^3$ & 2.10 \\ \hline
10 & U Her & 0.15 & 13. & $2.6 \cdot 10^{-7}$ & $1.1 \cdot 10^{-9}$ & 
0.42 & $1.1 \cdot 10^3$ & $3.2 \cdot 10^5$ & 0.06 \\ \hline
11 & FX Ser & 0.20 & 26. & $1.2 \cdot 10^{-5}$ & $1.7 \cdot 10^{-8}$ & 
0.14 & $8.1 \cdot 10^3$ & $8.4 \cdot 10^3$ & 0.98 \\ \hline
12 & NSV 11225 & 0.02 & 14. & $9.1 \cdot 10^{-6}$ & $1.1 \cdot 10^{-8}$ & 
0.12 & $1.2 \cdot 10^4$ & $2.8 \cdot 10^4$ & 0.65 \\ \hline
\end{tabular}
\end{center}
\normalsize

\caption[2]{Model estimates for sources in Table~1. Envelope optical
depth $\tau_{10}$, 
dust mass-loss $\dot M_{d}$, dust to gas ratio $\mu$, total luminosity $L_*$
and distance $d$ for each source are 
determined by model fitting parameters,
while total mass loss $\dot M$, total fluxes $F_{1 kpc}$ and gas outflow
velocities $v_e$ are derived by IR-radio-millimetric observations from 
Loup et al. 1993.}
\end{table}

Note that the dust to gas ratios are generally low ($\sim 10^{-3}$),
with the exception of WX Ser. We would like 
however to stress that estimates of $\mu$ 
are affected by uncertainties
in the grain average radius $a$; Eq.~2 shows that $\dot M_d$ is
proportional to $a$ (assuming a correct fitting of $\tau_{10}$), which 
is not, in this case, simply a normalization factor. The distances we
obtained are all in accordance with estimates by other authors (see
e.g. Loup et al. 1993).

Figure~5 shows the model spectra for a few sources reported in Table~1,  
superimposed to the IRAS LRS. 
All sources except those O-rich with less prominent silicate features 
(LML90's  ``Sil+'' and ``S'' subclasses) 
can be satisfactorily fitted by the available opacities. 
In the case of the S-star W And (``S'' subclass), to fit the LRS spectra
was necessary to use a mixture of silicate and carbonaceous
dust in the proportion 1:4, being impossible to model the source using
silicates dust only.
The models for the sources with weak silicate features in subclass
``Sil+'' were obtained by fitting the
continuum only; this allows us to produce  
reasonable values for the physical parameters listed in Table~2, but
is not able to reproduce the correct profile of the dust feature. 
In our opinion, this is probably related to an
inadequateness of the available 
silicate dust opacities for sources of LRS subclass ``Sil+'' and
``S'', but we discuss other possibilities in Section~5.

\begin{figure}
\centerline{\psfig{file=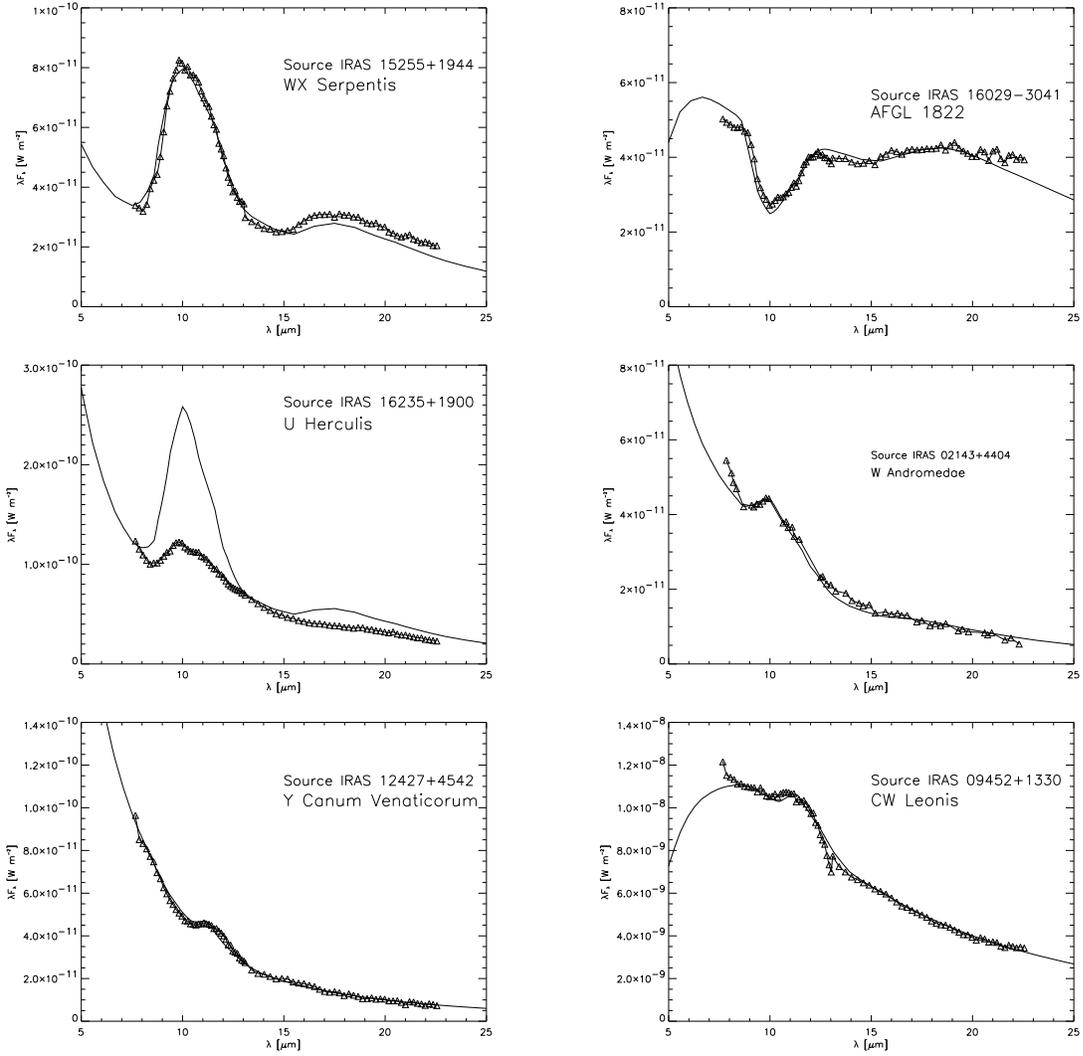,height=17cm}}
\caption[5]{Model spectra for 6 sources listed in Table~1. 
IRAS LRS data (triangles) are superimposed.}  
\end{figure}


\subsection{Temperature Profiles}

In spherical symmetry, the thermal structure of the model is expressed
by a radial law $T(r)$ giving the self-consistent dust radiative
equilibrium temperature for the dust grains. The temperature
profiles thus obtained depend on the dust chemical composition and 
on the   
optical depth. For large envelope radii they can be approximated by
a power law $T(r) \propto r^{-\alpha}$, where $\alpha$ is the
asymptotic value for the logarithmic derivative of the temperature
profile $T(r)$: 

$$\alpha = \lim_{r \to \infty} - {{d \log T} \over {d \log r}}
\eqno (3)$$

The inferred $\alpha $ values are restricted to a limited range (see
Table~3); however 
those for O-rich sources are systematically lower than those for C-rich 
ones (0.35 vs 0.40), while the case of the S-star modelled with a
mixed dust composition is intermediate.
This numerical result is  in good agreement
with the analitic formula derived by Harvey et al. (1991) for spherically
symmetric dust shells having power-law opacity $Q_\nu \propto
\lambda^{-\beta}$ and density $n(r) \propto r^{-2}$:

$$T(r) \propto r^{-2/(4+\beta)} \eqno (4)$$

For Oss1 silicates $\beta \simeq 2$ in the mid- and far-IR
wavelength range, while for AC1 Rouleau and Martin amorphous carbon 
is $\beta \simeq 1$ in the same spectral range; Eq.~4 would give
$\alpha \simeq 0.33$ for silicates and $\alpha \simeq 0.40$
for amorphous carbon, similar to our results,
appropriate for the outer parts of the envelopes, where the mid- and
far-IR thermal radiation is emitted.

\begin{table}
\begin{center}
\begin{tabular}{||l|c|c||} \hline 
Name & $\tau_{10}$ & $\alpha$ \\ \hline \hline
{\bf O-rich sources} & & \\ \hline 
R LMi & 0.15 & 0.348 \\ \hline
RT Vir & 0.15 & 0.349 \\ \hline
S CrB & 0.15 & 0.344 \\ \hline
U Her & 0.15 & 0.349 \\ \hline
WX Ser & 1.30 & 0.347 \\ \hline
AFGL 1822 & 8.00 & 0.332 \\ \hline \hline
{\bf S source} & & \\ \hline 
W And & 0.17 & 0.385 \\ \hline \hline
{\bf C-rich sources} & & \\ \hline 
Y CVn & 0.01 & 0.397 \\ \hline
X CnC & 0.15 & 0.400 \\ \hline
FX Ser & 0.20 & 0.404 \\ \hline
NSV 11225 & 0.20 & 0.404 \\ \hline
CW Leo & 1.30 & 0.404 \\ \hline
\end{tabular}
\end{center}

\caption[3]{Asymptotic values of the thermal structure power-law
exponent for the 
modelled sources, ordered for chemical type and increasing optical
depth $\tau_{10}$.}
\end{table}


\section{Discussion}

In order to account for the difficulties we found in modelling the O-rich
envelopes of subclass ``S'' and ``Sil+'', several possibilities can be
considered: deviation from spherical symmetry, mixing between O-rich
and C-rich dust and necessity of new opacities for silicates producing
LRS other than ``Sil'' subclass.

Evidences for non-spherical symmetry of AGB CSEs arise from radio
mapping, speckle interferometry, polarization studies and
spectroscopic observations 
as well as from the shapes of planetary nebulae (Balick 1993).
Departures from spherical symmetry should then be expected
also for our sources (see e.g. Silvestro et al. 1996 for a
photopolarimetric study of 5 sources in our sample), and may affect
even the thermal IR emission, as shown in Paper~I by the elongated
shape of CW Leo. Axisymmetric models of
AGB circumstellar envelopes have been recently developed by Collison \& 
Fix (1991) and Lopez et al. (1995); here basically the optical
depth depends on the orientation with respect to the line of
sight. These models thus introduce a new free parameter that allows one to
fit the dust feature strength; we notice however that the band
intensity remains unchanged with the slope of the continuum,
preventing the correct fitting of LRS spectra in our ``S'' and
``Sil+'' sources. Furthermore, it is not clear why deviations from
spherical symmetry should be important only for these subclasses of
envelopes. We conclude that, if a correct two-dimensional treatment of
the system geometry is essential for the spatial description of the IR
emissions, it does not significantly affect the spectra modelling.

In IE95 the IRAS colors of sources in the $2n$
class are reproduced admitting mixtures of silicates (plus 20\%
cristalline olivine) and amourphous carbon (in the ratio 1:4). The
possibility of such a mixing is controversial (dust formation in
chemical equilibrium models predicts the depletion of the less abundant
element between O and C, 
locked in CO molecules, see e.g. Salpeter 1974), but
cannot be completely ruled out, since numerous O-rich stars
display the SiC feature (see e.g. Willems \& de Jong, 1986 and Le Van
et al., 1992).
Even though this hypothesis is unattractive,
requiring the presence of mixtures only for sources of small optical
depth, the others being well fitted by ordinary silicate opacities, we
tested this prescription with our ``S'' and ``Sil+'' sources.

We have not obtained a real improvement in our ``Sil+'' model with mixed
dust, because the introduction of a new fitting parameter (the
silicate/carbon ratio) allowed only to reproduce the shape of the 9.8
$\mu$m feature, but not the slope of the continuum, even in the LRS
region. On the other hand, in the case of the ``S'' source, the injection
of a large quantity of amorphous carbon dust in the envelope (80\% of the
total) reduced the strenght of the silicate feature making
possible to fit the continuum in a large wavelengths
range (even thought the result of the fitting for the continuum is 
not as good as in the case of silicates
only). To understand this result, one should consider that the
S-stars like W And are considered objects in transition between the
O-rich and the C-rich class, and thus a contamination of silicate dust
in a C-rich circumstellar envelope is not at all surprising.

A third possibility to have weak silicate bands in O-rich
envelopes, as required in ``S'' and ``Sil+'' LRS, without the
introduction of carbonaceous
dust, is to assume annealing and aging of dust grains, as proposed by
Nuth \& Hecht (1990). Their grain condensation scheme 
starts with oxydation of SiO, AlO and OH in O-rich stellar winds,
producing grains characterized by weaker features and secondary
bands at 11-13 $\mu$m; fully grown silicates with strong features
condensate only later, and are thus present only in more evolved
O-rich envelopes. This view is in agreement with the 
classification schemes of LML90, assuming that LRS subclasses are
associated with dust of different ages. If the Nuth \&  
Hecht (1990) sequence is
correct, our O-rich sources with LRS subclass other than ``Sil''
should be associated with envelopes with less evolved dust. We have
not tested LML90 predictions, because
this would have required detailed dust opacities for each subclass,
presently not available. 
Fully developed radiative transfer modelling is in fact
necessary even to model optically thin envelopes as shown by IE95,  
and should be performed to extract dust opacities
from astronomical data.


\section{Conclusions}

Radiative transfer modelling of circumstellar envelopes, 
based on
simple assumptions as spherical symmetry and steady state outflows,
allows detailed fitting of infrared data for single sources and the
determination of system parameters as the distance, luminosity and
total amount of dust ejected by the central star. The correlation with
radio observations (giving informations of the total mass loss) will thus
provide estimates of the dust to gas mass ratio for each modelled source.
The accuracy of the fitting procedure depends mainly on the available
opacities, expecially in the mid-IR range where the dust features are
present; a good description of the mid-IR spectra is in fact essential
for the determination of the envelope optical depth and    
chemical composition. 

The necessity of accurate spectral
modelling is becoming more and more important taking into account 
the better spectroscopic capabilities of 
ground based mid-IR cameras and ISO data.

Our models confirm that the shape of carbonaceous dust features are
reproduced assuming a mixture of amorphous carbon, described by AC1
Rouleau \& Martin (1991) opacity, and (10\%, or less) P\'egouri\'e
(1988) SiC. Minor components (as PAH)
might be present, but the quality of IRAS
LRS do not allow their detection.

Models for O-rich envelopes associated to M and S stars suggest the
possibility of different varieties of silicate dust, even if mixing
between oxidic and carbonaceous dust should be considered, at least in
the case of transition objects as the S-stars. If the LML90 
scheme is valid, we can recognize in the
Ossenkopf~1 ``dirty silicates'' (OS92) the
correct opacity for subclass ``Sil'' silicates, associated to $2n$ LRS
spectra and a mixture of Ossenkopf~1 ``dirty silicates'' (OS92) with a
large amount of Rouleau \& Martin (1991) AC1 amorphous carbon (in the
ratio 1:4 for the subclass ``S'' sources).
For the other subclasses accurate fits need opacities 
which at present are not available.

The dust-to-gas mass ratios we obtained 
are generally low (between 0.1
and 0.5 \%), except in the case of the O-rich ``Sil'' source WX Ser
(1.5\%); there is a tendency for having lower dust-to-gas ratios 
for
C-rich envelopes, but this result may be influenced by uncertaintes in
the average grain radius (that can be different between oxidic and
carbonaceous dust) and by the inaccurate fitting of the silicate
features for O-rich sources other than ``Sil'' subclass. New models
taking into account the global dynamics of the system and coupling the
dusty component with the gaseous one will allow direct correlation
between IR and radio data, and will increase the accuracy in the
determination of the dust-to-gas ratio, giving better 
insight on the
dust condensation processes responsible for the mass loss in AGB stars
and the formation of circumstellar envelopes.




\begin{thebibliography}{}

\bibitem{}Balick, B. 1993, in: Planetary Nebulae, ed. R. Weinberger, A. 
Acker, Proc. IAU Symp. 155, Kluwer, Dordrecht, p. 131
\bibitem{}Boothroyd, A.I., Sackmann, I.J., Ahern, S.C. 1993, \apj 416 762
\bibitem{}Buss, R.H., Tielens, A.G.G.M., Snow, T.P. 1991, \apj 372 281
\bibitem{}Busso, M., Lambert, D.L., Beglio, L. 1995, \apj 446 775
\bibitem{}Busso, M., Origlia, L., Marengo, M., Persi, P., Ferrari-Toniolo,
M., Silvestro, G., Corcione, L., Tapia, M., Bohigas, J. 1996, \aa,
311 253
\bibitem{}Bussoletti, E., Colangeli, L., Borghesi, A., Orofino, V. 1987,
\aas 70 257
\bibitem{}Chan, S.J., Kwok, S. 1990, \aa 237 354
\bibitem{}Cherchneff, I., Barker, J.R. 1992, \apj 416 769
\bibitem{}Collison, A.J., Fix, J.D. 1991, \apj 368 545
\bibitem{}David, P., P\'egouri\'e 1995, \aa 293 833
\bibitem{}Draine, B.T. 1984, \apj 227 L71
\bibitem{}Draine, B.T., Lee, H.M. 1984 \apj 285 89
\bibitem{}Duley, W.W. 1993, in: Interstellar Dust, ed. L.J. Allamandola,
    A.G.G.M. Tielens, Proc. IAU Symp. 135, Kluwer, Dordrecht, p.141
\bibitem{}Frost, C.A., Lattanzio, J.C. 1996, in proc. ``Stellar
Evolution: What Should Be Done''; 32$^{nd}$ Li\'ege
Int. Astroph. Coll., 1995, in press
\bibitem{}Griffin, I.P. 1990, \mn 247 591
\bibitem{}Griffin, I.P. 1993, \mn 260 831
\bibitem{}Groenewegen, M.A.T. 1993, Ph.D. thesis, Univ. of Amsterdam
\bibitem{}Habing, H.J. 1996, \aar 7 97
\bibitem{}Hashimoto, O. 1995, \apj 442 286
\bibitem{}Harvey, P.M., Lester, D.F., Brock, D., Joy, M. 1991
\apj 368 558
\bibitem{}IRAS Catalogues and Atlases, Atlas of Low Resolution Spectra
1986, Iras Science Team, \aas 65 607
\bibitem{}Ivezi\'c, \v Z., Elitzur, M. 1995, \apj 445 415 (IE95) 
\bibitem{}Ivezi\'c, \v Z., Elitzur, M. 1996a,b, \mn in press
\bibitem{}Jones, A.P., Duley, W.W., Williams, D.A. 1990, \qjras 31 567
\bibitem{}Jones, T.W., Merrill, K.M. 1976, \apj 389 400
\bibitem{}Jura, M. 1996,\apj in press
\bibitem{}Justtanont, K., Tielens, A.G.G.M. 1992, \apj 389 400
\bibitem{}K\"aufl, H.U., Jouan, R., Lagage, P.O, Masse, P., Mestreau, P.,
Tarrius, A. 1992, The Messenger 70, 67
\bibitem{}Kr\"atschmer, W., Huffmann, D.R. 1978, \apss 61 195
\bibitem{}Little-Marenin, I.R., Little, S.J. 1990, \aj 99 1173 (LML90)
\bibitem{}Le Van, P.D., Sloan, G.C., Little-Marenin, I.R., Grasdalen,
G.L. 1992 \apj 392 702
\bibitem{}Lopez, B., M\'ekarnia, D., Lef\`evre, J. 1995 \aa 296 752
\bibitem{}Loup, C., Forveille, T., Omont, A., Paul, J.F. 1993,
\aas 99 291
\bibitem{}Martin, P.G., Rogers, C. 1987, \apj 322 374
\bibitem{}Mathis, J.S., Rumpl, W., Nordsieck, K.H. 1977, \apj 217 425
\bibitem{}Nuth, J.A.III, Hecht, J.H. 1990, \apss 163 79
\bibitem{}Omont, A., Loup, C., Forveille, T., te Lintel Hekkert,
P. Habing, H., Sivagnanam, P. 1993, \aa 267 515
\bibitem{}Ossenkopf, V., Henning, Th., Mathis, J.S. 1992 \aa 261
567 (OS92)
\bibitem{}P\'egouri\'e, B. 1988, \aa 194 335
\bibitem{}P\'egouri\'e, B., Papoular, R. 1985, \aa 142 451
\bibitem{}Puget, J.L., L\'eger, A. 1989, \araa 27 261
\bibitem{}Rouleau, F., Martin, P.G. 1991, \apj 377 526
\bibitem{}Rowan-Robinson, M. 1980, \apjs 44 403
\bibitem{}Salpeter, E.E. 1974, \apj 193 579
\bibitem{}Silvestro, G., Roseo, E., Marengo, M., Busso, B., Origlia, L.,
Scaltriti, F. 1996, in proc. I Torino Workshop: Evolution and
Nucleosynthesis in AGB Stars and Their Descendants, ed. M. Busso and
R. Gallino
\bibitem{}Sh\"onberner, D. 1996, comunication at the I Torino Workshop: 
Evolution and Nucleosynthesis in AGB Stars and Their Descendants
\bibitem{}Simpson, J.P. 1991, \apj 368 570
\bibitem{}Skinner, C.J., Whitmore, B. 1988, \mn 237 79
\bibitem{}Stencel, R.E., Nuth, J.A.III, Little-Marenin, I.R., Little,
S.J. 1990 \apj 350 L45
\bibitem{}Straniero, O., Gallino, R., Busso, M. et al. 1995,
\apj 440 L85
\bibitem{}van der Veen, W.E.C.J, Habing, H.J. 1988, \aa 194 125
\bibitem{}Vassiliadis, C., Wood, P.R. 1993, \apj 413 641
\bibitem{}Volk, K., Kwok, S. 1988, \apj 331 435
\bibitem{}Willems, F.J., de Jong, T. 1986, \apj 309 L39
\bibitem{}Wood, P.R., Bessel, M.S., Fox, M.W. 1983, \apj 272 99

\end{thebibliography}
\end{document}